# KANE-LIKE ELECTRONS IN TYPE II/III HETEROSTRUCTURES VERSUS DIRAC-LIKE ELECTRONS IN GRAPHENE

D. Dragoman – Univ. Bucharest, Physics Dept., P.O. Box MG-11, 077125 Bucharest, Romania, e-mail: danieladragoman@yahoo.com

**Abstract**

The propagation of charge carriers in graphene is compared to that in type II/III heterostructures for which a two-band Kane model is appropriate. In particular, conditions for a quantitative analogy between these two cases are searched for, and found to be quite restrictive. The analysis in this paper shows that the essential property of graphene is not the spinor character of its wavefunction but the linear dispersion relation, which does not hold in finite-gap two-band Kane-type semiconductors. Therefore, Kane-like and Dirac-like charge carriers behave differently, except in zero-bandgap semiconductor superlattices.

**Introduction**

Electron propagation in most semiconductors in the ballistic regime is described by the Schrödinger equation in which the electron rest mass $m_0$ is replaced by an effective mass. This description implies that the interaction between different energy bands, including the valence and conduction bands, can be neglected and that the electron wavefunction is scalar. On the other hand, in graphene the wavefunction of charge carriers is a two-component spinor, which satisfies a Dirac-like equation with vanishing effective mass; the two components of the spinor correspond to the contributions of the two triangular sublattices in the hexagonal crystalline structure (see the review in [1]). As a consequence, the charge carrier transport in graphene has distinct features compared to other semiconductors, which include the linear energy dispersion law, the absence of an energy bandgap and the impossibility of confining the charge carriers with electrostatic potentials. This difference persists even in the optical analogs of these two electron systems: the propagation of charge carriers in graphene is similar to that of polarization states of light [2], whereas the electron wavefunction in common semiconductors is analogous to one component of the electromagnetic field [3].

It seems therefore that a direct comparison of charge carrier transport in graphene and common semiconductors is impossible. However, we show in this paper that, under certain conditions, the transport of charge carriers in graphene is similar to that in type II (staggered gap) or type III (broken gap) heterojunctions [4], which can be described by a two-band Kane model [5]. The two-component wavefunction in these heterojunctions is found by solving a system of two coupled equations, similar to that describing the propagation of the electromagnetic field in a succession of directional couplers [6]. A comparison of Kane-like and Dirac-like electrons is helpful to understand the unique behavior of charge carriers in graphene. Our analysis emphasizes that the spinor character of wavefunction is not enough to render graphene a special status; the linear energy dispersion is as important as a two-component wavefunction. The understanding of this unique behavior of graphene is not trivial, and it is challenged lately by the discovery that Dirac-like fermions can exist also in other materials as long as the hexagonal lattice structure is preserved [7-9]. The availability of other structures with the same behavior of charge carriers as in graphene is important because graphene cannot presently be fabricated on large scale and with high quality. The results obtained in this paper show that it is not even necessary to have a hexagonal lattice in order to obtain charge carriers that obey a massless Dirac equation. Over a certain range of energies, the carrier transport, in particular the hole transport, in specific type II/III heterostructures described by a two-band Kane model mimics the transport of holes in graphene across interfaces between regions with different potential energies that are determined by the parameters of the materials forming the semiconductor heterostructure.

**Kane-like versus Dirac-like electrons at normal incidence**

Let us consider first a type II or type III heterojunction between two semiconductors labeled by $j = 1, 2$, with $x$ the stratification direction (see Fig. 1(a)). Inside each region $j$, at normal incidence ($q_y = 0$) the envelope wavefunctions in the conduction and valence bands, $\psi_{cj}$ and $\psi_{vj}$, respectively, satisfy the system of coupled equations [10,11]

$$\begin{pmatrix} E_{cj} - E & \hbar P_j q_{xj} \\ \hbar P_j q_{xj} & E_{vj} - E \end{pmatrix} \begin{pmatrix} \psi_{cj} \\ \psi_{vj} \end{pmatrix} = 0, \qquad (1)$$

where $E_{cj}$ and $E_{vj}$ are the band-edge energies of the conduction and valence bands, respectively, $P_j$ is the interband velocity matrix element between the conduction and (light-hole) valence bands, and $q_{xj}$ is the $x$-component of the wavevector $\mathbf{q}_j$ in semiconductor $j$.

The velocity matrix element is related to the effective mass $m_j$ in a semiconductor with energy gap $E_{gj} = E_{cj} - E_{vj}$ through

$$\frac{1}{m_j} = \frac{1}{m_0} + \frac{2P_j^2}{E_{gj}}. \qquad (2)$$

Equation (1) is similar to the Dirac-like equation satisfied in each region $j$ = 1, 2 by charge carriers in graphene, normally incident ($k_y$ = 0) on an interface between regions with potential energies $V_1$ and $V_2$ [1]:

$$\begin{pmatrix} V_j - E & \hbar v_F k_{xj} \\ \hbar v_F k_{xj} & V_j - E \end{pmatrix} \begin{pmatrix} \psi_{1j} \\ \psi_{2j} \end{pmatrix} = 0. \qquad (3)$$

In equation (3), $\psi_{1j}$, $\psi_{2j}$ are the two components of the spinor wavefunction in region $j$, $k_{xj}$ is the normal component of the wavevector $\mathbf{k}_j$ in graphene, and $v_F \cong c/300$ is the Fermi velocity in graphene. Different potential energies can be applied on different regions of a graphene flake through electrostatic gates, as shown in Fig. 1(b).

From (1) and (3) it follows that $P_j$ is equivalent to $v_F$, $q_{xj}$ is analogous to $k_{xj}$, and $V_j$ corresponds to $E_{cj} = E_{vj} = E_{0j}$. This last requirement implies that, at normal incidence, charge carriers in graphene in a region with potential energy $V_j$ behave analogous to charge carriers in a zero-bandgap semiconductor, in which the valence and conduction bands touch at a Dirac-like point $E_{0j}$. If the two-band Kane model is appropriate, the energy dispersion relation in such a semiconductor, $E = E_{0j} \pm |\hbar P_j k_{xj}|$, is linear, as in graphene, and the quantum wavefunction has the same form as in graphene. Zero-bandgap semiconductors that can be treated with the Kane model are HgTe, and long-period InAs/GaSb superlattices [12].

InAs/GaSb superlattices are commonly described by a two-band Kane model, while basic properties of HgTe and HgTe-based superlattices could be described by a two-band model [13-14] (more than two bands are taken into account in more refined models).

According to the considerations above, a similar propagation of charge carriers in graphene and semiconductor heterostructures at an interface implies that the velocity matrix element has the same value in all semiconductor layers (the Fermi velocity in graphene is the same, irrespective of the potential energy value), requirement that cannot be fulfilled unless we deal with a two-dimensional electron gas in a zero-bandgap semiconductor on which different gate potentials are applied. In this case, a perfect transmission of the electron wavefunction through the heterostructure at normal incidence, as in graphene, is not surprising since no quantum barrier layer can be identified for either electrons or holes. Moreover, a quantitative analogy between normally incident charge carriers in graphene and in a zero-bandgap semiconductor exists if

$$E_{02} = \frac{E_{01}(V_2 - E) + E(V_1 - V_2)}{V_1 - E}. \qquad (4)$$

However, a more general relation can be found if the analogies are carried on between the (normalized) equations

$$\begin{pmatrix} (E_{cj} - E)/\hbar P_j & q_{xj} \\ q_{xj} & (E_{vj} - E)/\hbar P_j \end{pmatrix} \begin{pmatrix} \psi_{cj} \\ \psi_{vj} \end{pmatrix} = 0, \quad \begin{pmatrix} (V_j - E)/\hbar v_F & k_{xj} \\ k_{xj} & (V_j - E)/\hbar v_F \end{pmatrix} \begin{pmatrix} \psi_{1j} \\ \psi_{2j} \end{pmatrix} = 0. \qquad (5)$$

In this case, although the analog of graphene is still a zero-bandgap Kane-type semiconductor with a linear dispersion relation around a Dirac-like point, the propagation of normally incident charge carriers in graphene can be mimicked by charge carriers propagating across an interface between zero-bandgap semiconductors if

$$E_{02} = \frac{E_{01}(V_2 - E)P_2 + E(V_1P_1 - V_2P_2)}{(V_1 - E)P_1}. \tag{6}$$

Although common type II/III heterostructures in which (1) applies, such as InAs/AlSb/GaSb [10, 11], involve semiconductors with a narrow but finite energy gap, so that a realization of graphene-like propagation of charge carriers in these heterojunctions is not obvious, long-period InAs/GaSb superlattices [12] can be envisaged as structures in which such an analogy could be observed.

**Kane-like versus Dirac-like electrons at oblique incidence on an interface**

An analogy of graphene with a more realistic semiconductor heterostructure, which consists of finite-bandgap semiconductor layers, can be found if we consider obliquely-incident charge carriers, for which $q_y \neq 0$. In this case, in each layer $j$ the envelope wavefunctions satisfy the following equation

$$\begin{pmatrix} E_{cj} - E - \dfrac{\hbar^2 \alpha^2 P_j^2 q_y^2}{E_{vj} - E} & \hbar P_j(q_{xj} - i\beta q_y) \\ \hbar P_j(q_{xj} + i\beta q_y) & E_{vj} - E \end{pmatrix} \begin{pmatrix} \psi_{cj} \\ \psi_{vj} \end{pmatrix} = 0. \tag{7}$$

Such an equation has been shown to describe the dynamics of charge carriers in a type II InAs/AlSb/GaSb/AlSb/InAs heterostructure, with $\alpha = \sqrt{3/2}$ and $\beta = -1/2$ [15]. The dispersion relation of electrons determined from (7) is

$$(E_{cj} - E)(E_{vj} - E) = \hbar^2 P_j^2[(\alpha^2 + \beta^2)q_y^2 + q_{xj}^2] \tag{8}$$

and the wavefunction can be expressed as

$$\begin{pmatrix} \psi_{cj} \\ \psi_{vj} \end{pmatrix} = \frac{1}{\sqrt{2}} \begin{pmatrix} 1 \\ p_j \exp(i\vartheta_j) \end{pmatrix} \exp(iq_{xj}x + iq_y y), \qquad (9)$$

where $p_j = \text{sgn}(E - E_{vj})$ and $\vartheta_j = \text{Arg}[(E - E_{vj})/\hbar P_j(k_{xj} - i\beta k_y)]$. Equation (7) is similar to that satisfied in each region $j$ by obliquely incident charge carriers in graphene (with $k_y \neq 0$) [1],

$$\begin{pmatrix} V_j - E & \hbar v_F (k_{xj} - ik_y) \\ \hbar v_F (k_{xj} + ik_y) & V_j - E \end{pmatrix} \begin{pmatrix} \psi_{1j} \\ \psi_{2j} \end{pmatrix} = 0, \qquad (10)$$

for which the dispersion relation can be written as

$$(E - V_j)^2 = \hbar^2 v_F^2 (k_{xj}^2 + k_y^2), \qquad (11)$$

whereas (9) has the same form as the wavefunction in graphene,

$$\begin{pmatrix} \psi_{1j} \\ \psi_{2j} \end{pmatrix} = \frac{1}{\sqrt{2}} \begin{pmatrix} 1 \\ s_j \exp(i\phi_j) \end{pmatrix} \exp(ik_{xj}x + ik_y y) \qquad (12)$$

where $s_j = \text{sgn}(E - V_j)$ and $\tan \phi_j = k_y / k_{xj}$. Note, however, that the dispersion relation in graphene is linear, whereas (8) is not: for $q_{xj} = q_y = 0$, the energy of charge carriers in the semiconductor has two possible solutions, $E = E_{cj}$ and $E = E_{vj}$, whereas for graphene $k_{xj} = k_y = 0$ corresponds to the Dirac point, for which $E = 0$ in the absence of an applied potential.

Despite this difference, a precise analogy between (7) and (10) can still be found if $V_j$ is equivalent to $E_{vj} = E_{cj} - \hbar^2 P_j^2 \alpha^2 q_y^2/(E_{vj} - E)$, i.e. if $q_y$ is chosen such that ($P_j$ is real)

$$q_y = \sqrt{E_{gj}(E_{vj} - E)} / |\hbar P_j \alpha|. \qquad (13)$$

Equation (13) implies that the propagation of charge carriers in type II/III heterojunctions mimics that in graphene if $E < E_{vj}$, i.e. if they are holes. For this reason, we focus in the following on analogies between hole propagation in graphene and semiconductor heterostructures. In addition, if only electrostatic potentials are applied, the wavevector in graphene is real, which implies that $q_{xj}$ should also be real; $q_{xj}$ is determined from (8) once $q_y$ is found from (13). It should be mentioned that, unlike in other materials, the electrical transport in graphene is ambipolar and electron and hole states have (near the Dirac point) the same mobitily; these states differ only by the sign of energy (positive and negative, respectively). Therefore, there is no difference between performances of devices based on electron or hole transport in graphene.

In the following, as in Ref. 3, we design a type II/III semiconductor heterostructure that has the same transmission/reflection and traversal time characteristics for holes as a gated region in graphene, on which only electrostatic potentials are applied. For both graphene and heterostructure cases the boundary conditions are the same since the envelope wavefunctions, as well as the spinor components in graphene must be continuous at the interface.

The first example is that of an interface between regions labeled with $j = 1$ and $j = 2$ with potential energies $V_j$ in graphene, and, respectively, with parameters $E_{cj}$, $E_{vj}$, $P_j$ (or $m_j$) in a semiconductor heterostructure. The intention is to establish the relations between these parameters for which we obtain the same reflection coefficient at the interface in both cases. The reflection coefficient is given by

$$r = \frac{\exp(i\varphi_1) - \exp(i\varphi_2)}{\exp(-i\varphi_1) + \exp(i\varphi_2)} \qquad (14)$$

where $\varphi_j$, $j = 1, 2$, denotes $\phi_j$ for the graphene case and $\vartheta_j$ for the heterostructure. The reflection probability of charge carriers at the interface is $R = |r|^2$, and the transmission probability is $T = 1 - R$. Note that for normal incidence $\varphi_1 = \varphi_2 = 0$, and all charge carriers are transmitted across the interface. Because the reflection coefficient depends only on $\varphi_j$, the same reflection coefficient for the graphene and semiconductor heterostructure cases is obtained only if

$$\frac{E - E_{v1}}{\hbar P_1(q_{x1} - i\beta q_y)} = \frac{E - V_1}{\hbar v_F(k_{x1} - ik_y)}, \qquad \frac{E - E_{v2}}{\hbar P_2(q_{x2} - i\beta q_y)} = \frac{E - V_2}{\hbar v_F(k_{x2} - ik_y)}. \qquad (15)$$

Moreover, one must be sure that $q_y$ remains the same in the two semiconductor layers, requirement that limits the graphene/heterostructure analogy to only one energy value, determined from (13) as

$$E = \frac{E_{v1} P_2^2 E_{g1} - E_{v2} P_1^2 E_{g2}}{P_2^2 E_{g1} - P_1^2 E_{g2}}. \qquad (16)$$

For this value of $E$, we must be sure that $q_y$, $q_{x1}$ and $q_{x2}$, found from (13) and (8), are all real. This is quite a challenge since this condition is not satisfied in several common type II/III heterostructures, for example in Si/Ge or in InAs/AlSb/GaSb heterostructures, but is satisfied in InP/GaAsSb. The energy band diagram of the latter heterostructure, given in Ref. 16, is illustrated in Fig. 2(a).

Once $q_y$, $q_{x1}$ and $q_{x2}$, and hence $\vartheta_1$ and $\vartheta_2$ are determined, the potential energies $V_1$ and $V_2$ are determined from the equations

$$E - V_{1,2} = \frac{k_y \hbar v_F}{\sin \phi_{1,2}}, \tag{17}$$

with $E$ the solution of (16) and the identifications $k_y = \beta q_y$, $\vartheta_1 = \varphi_1 = \phi_1$ and $\vartheta_2 = \varphi_2 = \phi_2$. So, a charge carrier with energy $E$ propagates across a heterostructure in the same way that across an interface between two regions in graphene with potential energies $V_1$ and $V_2$ given by (17). In particular, the reflection coefficient of charge carriers is the same in the two cases.

Of course, it would be interesting if this analogy could be extended to a whole range of energies, and not just for one energy value. This is possible only if the bandgap of one or both semiconductor layers can be changed. Then, the energy range for the analogy to hold follows from (16) and is determined by the variation range of $E_{gj}$. Doping or temperature variations are known mechanisms of bandgap modification. For instance, if we consider the second mechanism in the InP/GaAsSb heterostructure and consider the temperature dependence of $E_{gj}$ in the two semiconductors given in Ref. 17 (for GaAsSb a mean variation was considered between that of GaAs and GaSb), the reflection probability of electron waves decreases with temperature, as can be seen from Fig. 2(b). The effective masses of light holes are taken from Ref.17, the semiconductor labeled with 1 (2) being InP (GaAsSb). The temperature-dependent reflection probability in Fig. 2(b) at the interface between InP and GaAsSb is the same as that at an interface between two regions in graphene with potential energies $V_1$ and $V_2$, if these applied potential energies depend on temperature as in Fig. 3. Note that the potential energies are negative, situation that corresponds to hole propagation.

Both $R$ and $V_1$ and $V_2$ have a small range of variation since the bandgaps of InP and GaAsSb have a weak temperature dependence.

The temperature is in this case an additional parameter needed to extend the graphene/semiconductor analogy for a range of energy values. An additional parameter (additional dimension of the system or the phase of a phase modulator) was also needed to establish the classical optics/semiconductor heterostructure analogies in Ref. 3. Similar to the situation in Ref. 3, in our case this additional parameter is required by the difference in the dispersion relations in graphene and Kane-type heterostructures.

A second problem is to find an InP/GaAsSb/InP heterostructure with an identical transmission coefficient through a finite-width GaAsSb layer as that through a region with potential energy $V_2$ in graphene sandwiched between identical semi-infinite regions with potential energies $V_1$. More precisely, the transmission coefficient through such a layer with thickness $L$ is given by

$$t_l = \frac{2i \exp[i(\gamma_{x2} - \gamma_{x1})L]\cos\varphi_1 \cos\varphi_2}{1 + \cos(\varphi_1 + \varphi_2) + \exp(2i\gamma_{x2}L)[\cos(\varphi_1 - \varphi_2) - 1]}, \tag{18}$$

the probability of charge carrier transmission through layer 2 being $T_l = |t_l|^2$, while the reflection probability is $R_l = 1 - T_l$. Here $\gamma_{x1,2}$ stands for $k_{x1,2}$ in the case of graphene, and for $q_{x1,2}$ for the semiconductor heterostructure. Again, for normal incidence or $L \to 0$, we obtain $T_l = 1$.

Because the probability of hole transmission determined from (18) depends not only on $\varphi_1$ and $\varphi_2$ but also on $\gamma_{x2}L$, we must impose an additional condition for the propagation of charge carriers in graphene and semiconductor heterostructures in order to obtain the same $T_l$ in both cases: $q_{x2}L_s = k_{x2}L_g$, where $L_s$, $L_g$ are the respective widths of layer 2 in semiconductor and graphene. If $L_s$ and $L_g$ are fixed, and $q_{x2}$, $\varphi_1 = \phi_1$ and $\varphi_2 = \phi_2$ are determined as above, the potential energy in region 2 in graphene is determined from

$$k_{x2} = \frac{E - V_2}{\hbar v_F} \cos\phi_2 = \frac{L_s}{L_g} q_{x2}, \tag{19}$$

and $V_1$ is determined subsequently from the equality condition for the tangential component of the wavevector, expressed as

$$(E - V_1)\sin\phi_1 = (E - V_2)\sin\phi_2. \tag{20}$$

Simulations for the same heterostructure as above show that the transmission probability of holes through a $L_s = 5$ nm wide GaAsSb layer is the same as through a graphene region of width $L_g = 50$ nm, if the potential energies in this region, $V_2$, and in the regions that surround it, $V_1$, depend on the temperature of the InP/GaAsSb heterostructure as shown in Fig. 4. Note that the temperature dependence of the potential energies in graphene, for which $T_l$ is the same as in a InP/GaAsSb/InP heterostructure, is different from that in Fig. 3. The reason is that the potential energies, although also negative, are determined in this case from different conditions.

The temperature dependence of $T_l$ for the InP/GaAsSb/InP heterostructure is represented in Fig. 5 with solid line. With dashed line we have represented the temperature dependence of the traversal time, defined in terms of the group velocity $v_g = J/\rho$ in layer 2 (GaAsSb) as

$$\tau = \int \frac{dx}{v_g(x)} = \int \frac{\rho(x)dx}{J}, \tag{21}$$

where $\rho = |\psi_{c2}|^2 + |\psi_{v2}|^2$ is the probability density and $J = P(\psi_{c2}\psi_{v2}^* + \psi_{c2}^*\psi_{v2})$ is the probability current along $x$ [18]; similar curves hold for graphene if the potential energies are related to temperature as in Fig. 4. An interesting property of Fig. 5 is that at a certain temperature/wavevector component tangent to the interface the transmission probability equals 1. This is indeed a feature encountered for charge carrier propagation through a gated region in graphene [19], where the transmission probability equals 1 not only for normal incidence but also for certain angles of incidence. The simulations in Fig. 5 confirm, therefore, that hole propagation in InP/GaAsSb/InP heterostructures under the precise conditions defined above mimic indeed hole propagation in graphene. At the same time, $\tau$ does not show a maximum where the transmission probability is maximum, revealing that the maximum in $T_l$ is not of a resonant nature, as in type II superlattices composed of barriers and wells [18].

**Conclusions**

It was shown that carrier propagation in graphene can be mimicked by the propagation of holes in type II/III heterostructures under very special conditions. In particular, a quantitative analogy between these two cases is valid for a single value of energy, unless the parameters of semiconductors, in particular their bandgap, can be modified. The graphene/semiconductor heterostructure comparison detailed in this paper shows that it is not enough to have a system of coupled equations for quantum wavefunctions in order to obtain a graphene-like behavior. The essential property of graphene is not the spinor character of its wavefunction but the linear dispersion relation, which does not hold in finite-gap two-band Kane-type semiconductors. This is the reason why Kane-like and Dirac-like charge carriers behave differently, unless zero-bandgap semiconductor superlattices are considered.

**Figure captions**

Fig. 1  Charge carriers in a type II/III heterojunction between semiconductors 1 and 2 (a) obey a similar propagation law as across an interface between regions with different potential energies in graphene, induced via electrostatic gates (b).

Fig. 2  (a) Schematic band diagram of the InP/GaAsSb interface, and (b) the corresponding reflection probability dependence on temperature.

Fig. 3  Temperature dependence of potential energies in graphene regions for which $R$ is the same as in Fig. 1(b).

Fig. 4  Temperature dependence of potential energies in graphene regions, for which hole propagation through a 50 nm wide region is similar to propagation through a 5 nm GaAsSb layer sandwiched between InP regions.

Fig. 5  Temperature dependence of transmission probability $T$ (solid line) and of traversal time $\tau$ (dashed line) for the InP/GaAsSb/InP heterostructure with a 5-nm-wide GaAsSb layer.

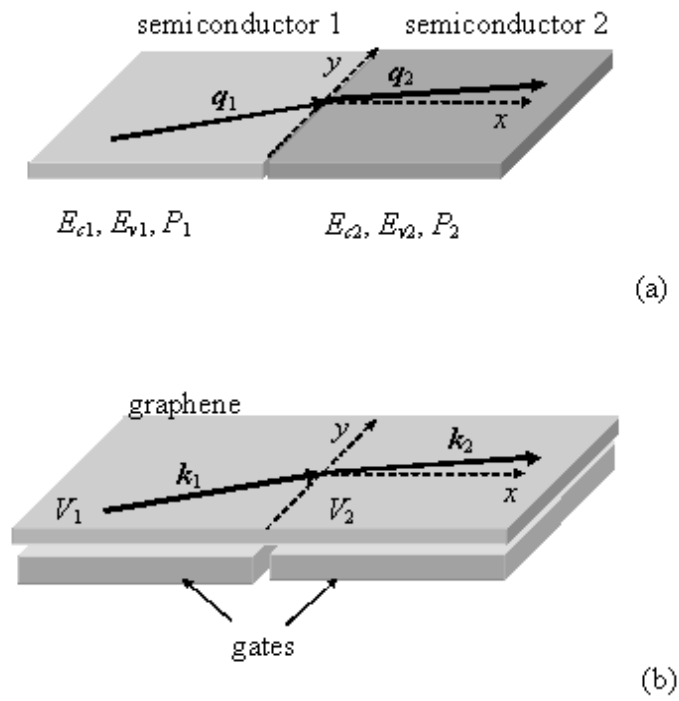

Fig. 1

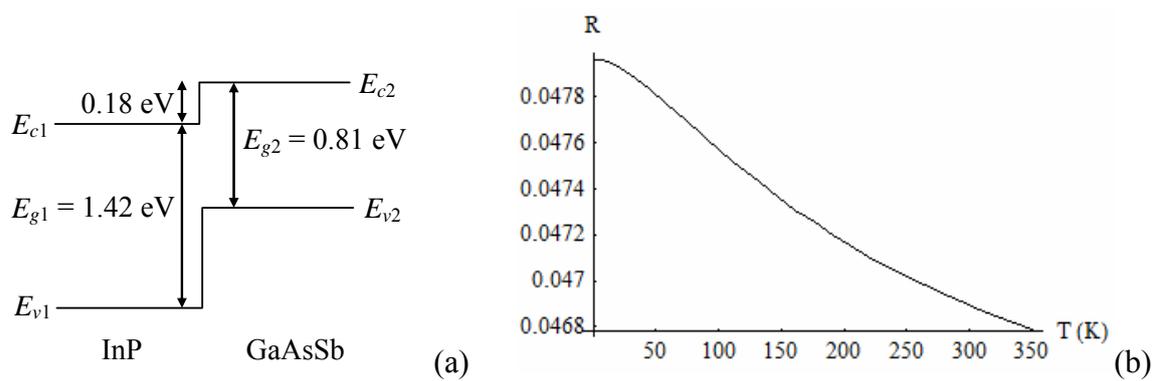

Fig. 2

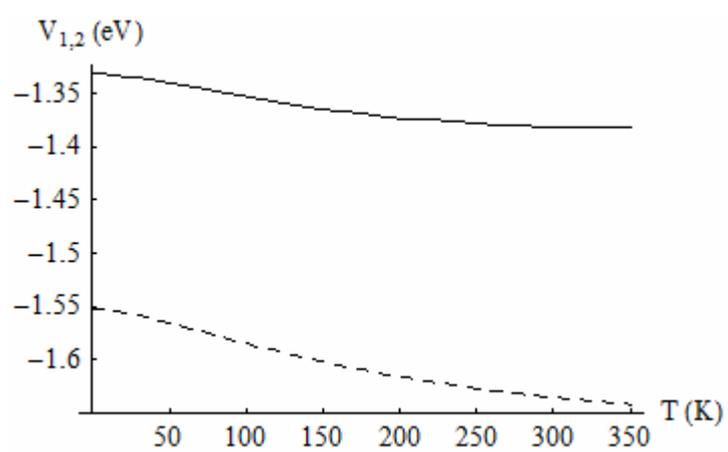

Fig. 3

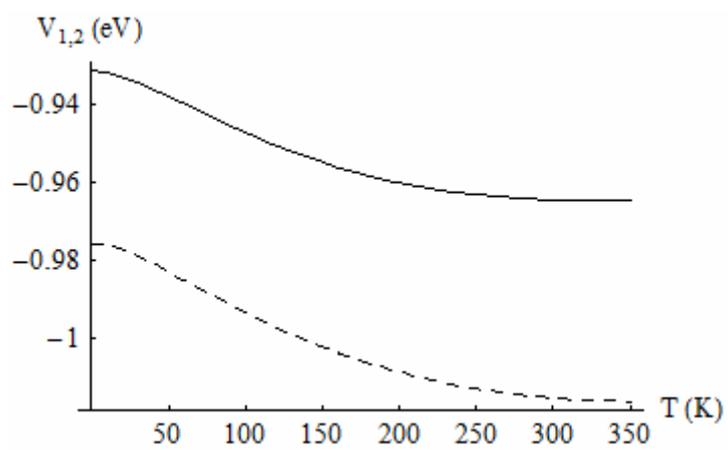

Fig. 4

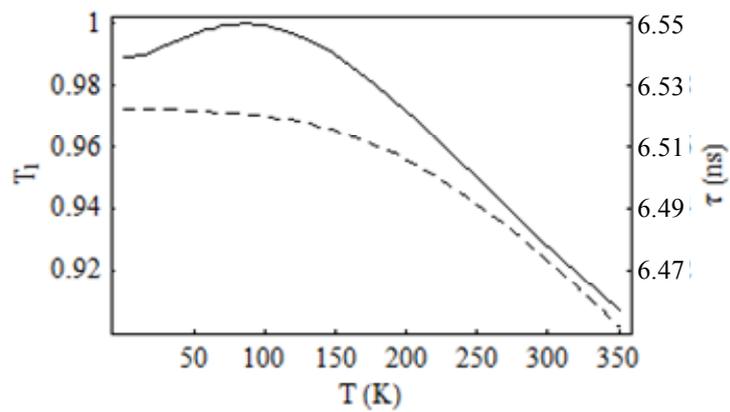

Fig. 5